\documentclass[english]{elsart}
\usepackage[T1]{fontenc}
\usepackage[latin1]{inputenc}
\usepackage{amsmath}
\usepackage{setspace}
\usepackage{amssymb}

\makeatletter

\providecommand{\LyX}{L\kern-.1667em\lower.25em\hbox{Y}\kern-.125emX\@}

\usepackage{setspace}

\usepackage[numbers]{natbib}

\usepackage[dvips]{graphicx}
\usepackage{epic,epsfig}
\usepackage{subfigure,float}
\usepackage{amsmath,amsfonts,amssymb}
\usepackage{inputenc}
\usepackage{natbib}
\usepackage{euscript}

\makeatother
\begin{document}
\begin{frontmatter}

\title{Scale-Free Networks Generated By Random Walkers}
\author{Jari Saram\"{a}ki\corauthref{cor1}}, and
\ead{jsaramak@lce.hut.fi}
\author{Kimmo Kaski}
\corauth[cor1]{Corresponding author.}

\address{Laboratory of Computational Engineering, Helsinki University of Technology, 
P.O. Box 9203, FIN-02015 HUT, Finland}

\begin{abstract}
We present a simple mechanism for generating undirected 
scale-free networks using
random walkers, where the network growth is determined by choosing parent
vertices by sequential random walks. We show that this mechanism produces
scale-free networks with degree exponent $\gamma = 3$ and clustering
coefficients depending on random walk length.
 The mechanism can be interpreted in terms
of preferential attachment without explicit knowledge of node degrees.
\end{abstract}

\end{frontmatter}

Since the seminal paper by Watts and Strogatz~\cite{SW}, much research has 
focused on the properties of small-world and scale-free 
networks~\cite{Strogatz:exploring,Albert:review,Dorogovtsev:evolution},
as these have been found to resemble many naturally occurring networks, such
as social networks~\cite{Amaral:social,Girvan:Newman}, scientific collaboration
networks~\cite{Newman:science}, the WWW~\cite{Albert:www,Huberman:www} and
metabolic networks~\cite{Jeong:metabolic}. Most naturally occurring networks
can be characterized by a 
high degree of clustering together with small average node-to-node
distance~\cite{SW}. In addition, these networks often display a power-law
degree distribution $p(k) \propto k^{-\gamma}$, where $k$ denotes 
vertex degree and $\gamma$ the degree exponent, which is typically observed
to lie within the range $\gamma\sim 2-3$.
The emergent power law
was originally
explained by Barab\'{a}si and Albert (BA) in terms of combining
network growth and preferential attachment ~\cite{BA}.
 The BA
preferential attachment model simply states that when a new node is 
added to the network, it is preferentially linked
 to nodes already possessing a large 
number of links. This intuitive mechanism results in scale-free 
networks with $\gamma=3$. 

Over the recent years, several new models for generating scale-free networks
have been
proposed~\cite{Vazquez2000,Holme2001,Dorogovtsev2001a,Dorogovtsev2002a,
Klemm2002a,Klemm2002b,Ravasz2003,
Vazquez2003}. One motivation has been to better capture the clustering
properties of real-world networks~\cite{Holme2001,Klemm2002a,Klemm2002b,
Ravasz2003}, as
these tend to exhibit far larger degree of clustering than BA networks.
Furthermore, the original BA model does not explicitly state 
how the preferential attachment comes to be. The algorithm
utilizes global knowledge of vertex degrees but in the case of real-world networks such as social networks or the WWW, new ``vertices'' joining these networks
rarely have such knowledge. Despite of that, power laws emerge. 
Thus, determining growth-directing rules
which only utilize local information on vertex degrees and connections
is of importance. These ideas have been elaborated in 
Refs.~\cite{Vazquez2000,Vazquez2003}, where it was shown that in directed
networks, random walkers moving along the edges of a network 
and probabilistically
creating new links to their current locations eventually 
results in scale-free degree
distributions. The idea
of reaching highly connected vertices by following random links was also
utilized by Cohen et al.~\cite{Cohen:immunization} in the
context of effective immunization strategies. 

Here, we present a simple undirected
network growth mechanism based on random walks of fixed length,
and show that it leads to a scale-free network structure with
BA degree exponent $\gamma = 3$. Furthermore, the clustering properties
of the network are determined by the random walk length. The algorithm
 goes as follows:

(i) The network is initialized with $m_0$ vertices, connected to each other.

(ii) A random vertex is chosen as the starting point of the random walk.

(iii) At each step of the walk, the walker moves to a randomly chosen
neighbor of the current vertex. After $l$ steps,
the vertex at which the random walker has arrived is marked. The walk
is repeatedly continued for $l$ steps until
$m \leq m_0$ different vertices are marked. 

(iv) A new vertex is added to the network and connected to the $m$ marked 
vertices by undirected links,
and the whole process is repeated starting from step (ii),
until the network has grown to the desired size of $N$ vertices.

Note that in step (iii), the walker is allowed to trace its steps backwards.
Thus the walk is not self-avoiding, as it would otherwise easily get stuck.
If the walker arrives at an already marked vertex after $l$ steps, 
a new $l$-length walk is started from that vertex.  

In this algorithm, preferential attachment follows from the fact that the 
probability of the random walker ending up at a highly connected vertex is 
higher than that of it ending up at a vertex with less connections. 
In the context of e.g. WWW or social networks, the idea is intuitively 
appealing. Analogously, we tend to learn to know new people through
those people we already know, which quickly leads us to ``popular'' persons
without explicitly looking for them.

In the following, we show that choosing vertices by the 
random walk method is equivalent to
the BA preferential attacment rule~\cite{BA}, 
stating that the probability $P$ of
node $i$ being chosen depends on its degree $k_i$:
\begin{equation}
P(i)=\frac{k_i}{\sum_j{k_j}}. \label{eq1}
\end{equation}
Here $\sum_j{k_j}$ denotes the sum over the degrees of all nodes, that is,
the total number of connections within the network.
Let us define $P(A)$ as the probability that vertex $A$ is chosen as the 
initial vertex, $P(A)=1/N$, and $P(B)$ as the probability that we arrive
at its neighbor vertex $B$ by following one of the $k_A$ links 
attached to $A$.
 We can derive an
expression for
$P(B)$ by utilizing the Bayes rule:
\begin{equation}
P(B)=\frac{P(B|A)P(A)}{P(A|B)}, \label{eq2}
\end{equation}
where the conditional 
probability $P(B|A)$ denotes the probability of arriving
 at $B$ if $A$ is chosen as the starting vertex. This probability
 equals the probability that of 
the $k_A$ links out of node $A$, the correct link is followed:
\begin{equation}
P(B|A)=\frac{1}{k_A}. \label{eq3}
\end{equation}
Likewise, the conditional
probability $P(A|B)$, i.e. the probability that if the walk arrived at $B$ it
originated at $A$, can be written as 
\begin{equation}
P(A|B)=\frac{1}{k_B}. \label{eq4}
\end{equation}
Combining all the above we arrive at
\begin{equation}
P(B)=\frac{k_B}{Nk_A}. \label{eq5}
\end{equation}
If we now continue the walk and utilize the derived result for $P(B)$
in calculating the probability of the walker
ending up at node $C$ within one single
step, we get
\begin{equation} 
P(C)=\frac{k_C}{Nk_A}, \label{eq6}
\end{equation}
as $k_B$ cancels out. Thus, the random walk length $l$ 
does not influence the probabilities of nodes to be chosen at all, 
although it plays a role in determining the clustering
properties of the network, as we shall see below. 
 Finally, as node $A$ was chosen randomly, its degree $k_A$ equals,
on the average, the average degree $\langle{k}\rangle$ of the network,
 and since 
$Nk_A=N\langle{k}\rangle=\sum_j{k_j}$, 
we arrive at the BA preferencial attachment rule (1), with the
well known consequence for the degree distribution $p(k)\propto k^{-3}$,
i.e. the degree exponent $\gamma=3$. It should be noted that 
degree-degree correlations could in principle 
influence this outcome; however,
no significant correlations were found in simulated 
networks generated by our random walk
method.

Figure 1 displays the degree distribution $p(k)$ versus $k$ on log-log scale, 
obtained by simulations for
 $m=2,4,8,16$, with $N=10^6$, and averaged over 100 runs each.
The random walk length was chosen $l=1$. In this figure, 
the solid lines indicate
corresponding BA degree distributions of the
form~\cite{Dorogovtsev2000a,Krapivsky2000,Albert:review},
\begin{equation}
p(k)=\frac{2m(m+1)}{k(k+1)(k+2)}, \label{BAdistr}
\end{equation} 
which at the limit $m\gg 1$, $k \gg 1$ can be written in the common form
$p(k)=2m^2/k^3$. It is evident that 
the simulated degree distributions match very well with
the theoretical ones. We have also repeated the same runs
for larger values of $l$ and obtained very similar results, confirming
the fact that the random walk length does not seem to
influence the degree distribution.

However, as mentioned above, the random walk length $l$ is not irrelevant
for the network properties. When $m>1$, one can expect that 
short random walk lengths give rise to highly clustered networks. 
The degree of clustering should be especially high when $l=1$, 
as every growth step then results in the formation
of at least 
$m-1$ triads of connected vertices.
 Clustering is measured in terms of the clustering coefficient, which 
denotes to what degree the neighbors  
of a vertex are also neighbors to each other,
and it has maximum value $C_{max}=1$ for a fully connected network.
The average clustering coefficient $C$ of the whole network~\cite{SW}
is defined as
the ratio of the existing links $E_i$ between neighbors of vertex $i$
to
the possible number of such connections, averaged over the whole network:
\begin{equation}
C=\langle{C_i}\rangle=\left\langle{\frac{2E_i}{k_i(k_i-1)}}\right\rangle.
\label{ceff_eq}
\end{equation}

The case $m=2$ is especially interesting: 
one can easily see that with $l=1$, the random walk model is 
identical to the Holme-Kim model ~\cite{Holme2001} with triad 
formation probability $P_t=1$. Furthermore, the clustering properties are
similar to the Dorogovtsev-Mendes-Samukhin 
model~\cite{Dorogovtsev2001a}, as well as to the  
highly clustered Klemm-Egu\'{i}luz model~\cite{Klemm2002b}. 
In this case, a new triad is formed every time a new vertex
is added to the network. We can calculate the clustering coefficient 
using a slight variation of the rate equation approach~\cite{Szabo2003}: 
for each vertex ${\partial E(k)}/{\partial k}=1$
with initial condition $E(2)=1$, resulting in $E(k)=k-1$. Thus,
the clustering coefficient of a vertex of degree $k$ equals $C(k)=2/k$.
Now the average clustering coefficient is obtained by using   
the degree distribution of Eq.~(\ref{BAdistr}):
\begin{equation}
C_{m=2,l=1}=\sum_{k=2}^{\infty}\frac{2}{k}\frac{2m(m+1)}{k(k+1)(k+2)},\label{C1}
\end{equation}
yielding a numerical value of $C_{m=2,l=1}\approx 0.74$.   
For $l=3$, one
can easily see that new triads are most likely formed 
by the walker returning
to the previous vertex at steps 2 or 3. The probabilities of both the above
cases can be
approximated to be $1/\langle{k}\rangle$, thus, on the average, 
$\partial{E}/\partial{k}\approx 2/\langle{k}\rangle=1/2$. 
Setting $E(2)=1/2$ we get
$C(k) \approx 1/k$, yielding $C_{m=2,l=3}\approx 0.37$. 
For larger odd values of $l$, the probability of the walker returning
to one of the neighbors of the first marked node decreases, causing
the clustering coefficient to decrease as well.
It is  interesting to note that a $C(k)\sim k^{-1}$-dependency similar to the
$l=1$ and $l=3$ cases
has been observed in several real-world networks~\cite{Ravasz2003}. 
The clustering coefficient distribution $C(k)=2/k$ of the $l=1$ case
 is also similar to the deterministic scale-free graph
of Ref.~\cite{Dorogovtsev2002a}.

Estimating the clustering coefficient values for even random walk lengths
turns out to be more difficult. In the $l=2$ case, 
$\partial{E}/\partial{k} \sim E/(k\langle k\rangle)$, i.e. a triad is formed
only if 
the walker at step 2 follows one of the already existing 
$E$ neighbor-neighbor links out of
$\sim k\langle k\rangle$ possible links. This equation suggests that for
(low) 
even values of $l$, the triad formation probability is significantly lower
than for odd values of $l$. It is also influenced by the initial conditions
(for $l=2$, if there are no clusters in the initial configuration, no triads
can be formed at all), and the configuration of links added at the
initial stages of growth. Unlike for odd $l$-values, one can 
expect an increase in clustering with increasing even $l$, as the probability
of the walker returning to the vicinity of its starting point increases.

Figure 2 illustrates the dependence of $C$ on $l$ for $m=2$ and 4,
 calculated by averaging over 100 runs of networks with size
$N=25,000$. The $l=1$ and $l=3$ values for the $m=2$ case match with the
above predictions, as does the overall shape of the curves.
At low enough values of $l$, the clustering
coefficient depends heavily on the random walk length. 
At odd values of $l$ the generated networks are highly clustered,
but there is a large difference   
between even and odd values of $l$. 
This difference disappears for large $l$, while at the same time
the clustering coefficient converges to a
constant value. The probable reason for this is that $l$ becomes
large enough for the $m$ nodes to be uncorrelated, and thus increasing it 
further has no more an effect. Furthermore, investigation 
of the clustering coefficients of networks generated in individual runs
reveals that when $l$ is small, there is only little scatter
for odd values of $l$, whereas the coefficients can differ greatly for
even values of $l$.


Finally, we have investigated the dependency of $C$ on the network size.
 Figure 3 illustrates $C$ as function of $N$
 for $m=4$ and $8$, and for several values of $l$. When
the random walks are short, $C$ appears quite independent of $N$. As discussed above, increasing (odd) $l$ results in a decrease in clustering. However, 
 as $l$ becomes large enough, we obtain clustering
similar to that of BA networks, where 
$C(N)\propto (\ln N)^2/N$ for $m\gg1$ ~\cite{Klemm2002b,Szabo2003}. 
The solid lines indicate
clustering coefficients $C(N)$ calculated by using 
the asymptotically similar formula (17) of Ref. \cite{Fronczak2003}, 
which is more accurate at low $m$. The theoretical curves show  
good agreement with our data. Thus, the BA
preferential attachment model could be considered as the limiting case
 of our model for $l\gg 1$.

In conclusion, we have presented a random-walk-based 
undirected network growth mechanism which results 
in scale-free networks with degree exponent $\gamma=3$ without utilizing
any knowledge of node degrees. Furthermore, we have analytically shown that
this mechanism leads to the Barab\'{a}si-Albert preferential attachment rule
for any random walk length, and thus, somewhat surprisingly,
even one-step random walks produce scale-free degree distributions. 
This clearly illustrates that growing networks can 
self-organize into scale-free structures even when based on simple, local
growth rules. We have also shown that the clustering degree 
of the network depends on the random walk length in addition to
the network size. This degree becomes similar to that of BA networks,
when the random walks are long enough.

\textbf{Acknowledgments}

This work is supported by the Academy of Finland, project No. 1169043 (Finnish
Centre of Excellence programme 2000-2005). 

\bibliographystyle{elsart-num.bst}
\bibliography{saramaki.bib}

\clearpage
\begin{figure}
\begin{center}
\includegraphics[width=\textwidth]{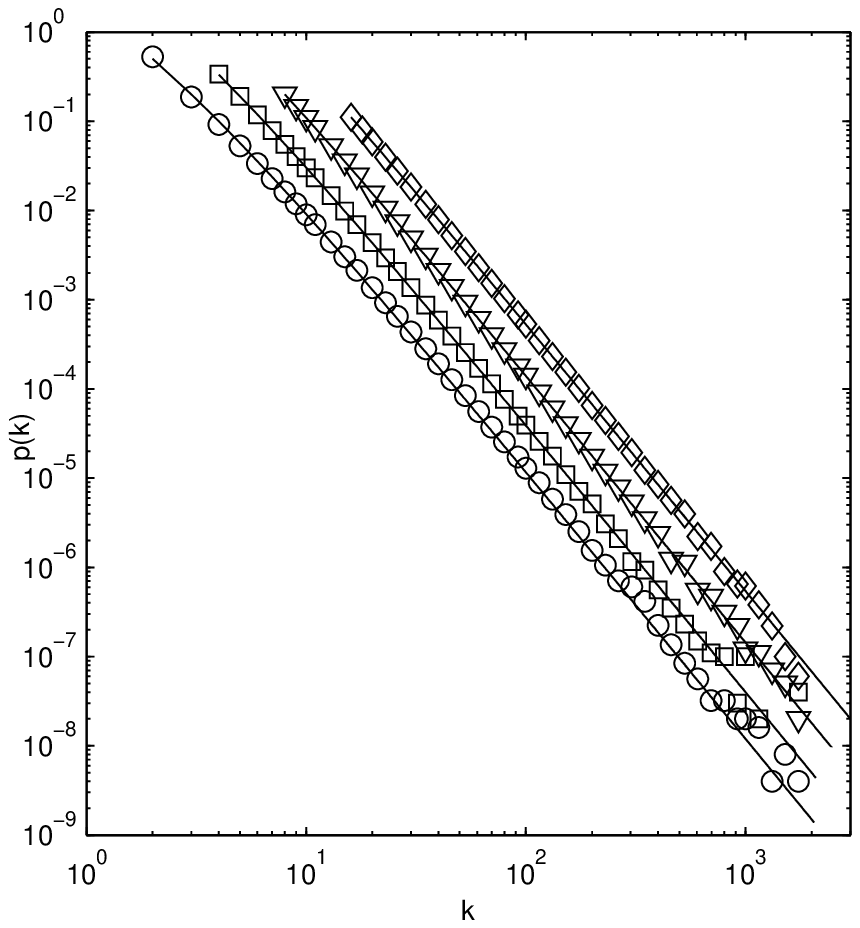}
\end{center}
\caption{Results of simulated degree distributions $p(k)$ as function of $k$,
calculated by averaging over 100 runs with networks grown to size
$N=10^6$, for $m=2$ ($\circ$), $m=4$ ($\square$), $m=8$ ($\triangledown$) and
$m=16$ ($\lozenge$). The networks were generated using random 
walks of length $l=1$. The solid lines indicate respective 
Barab\'{a}si-Albert degree distributions.} \label{fig1}
\end{figure}

\clearpage
\begin{figure}
\begin{center}
\includegraphics[width=\textwidth]{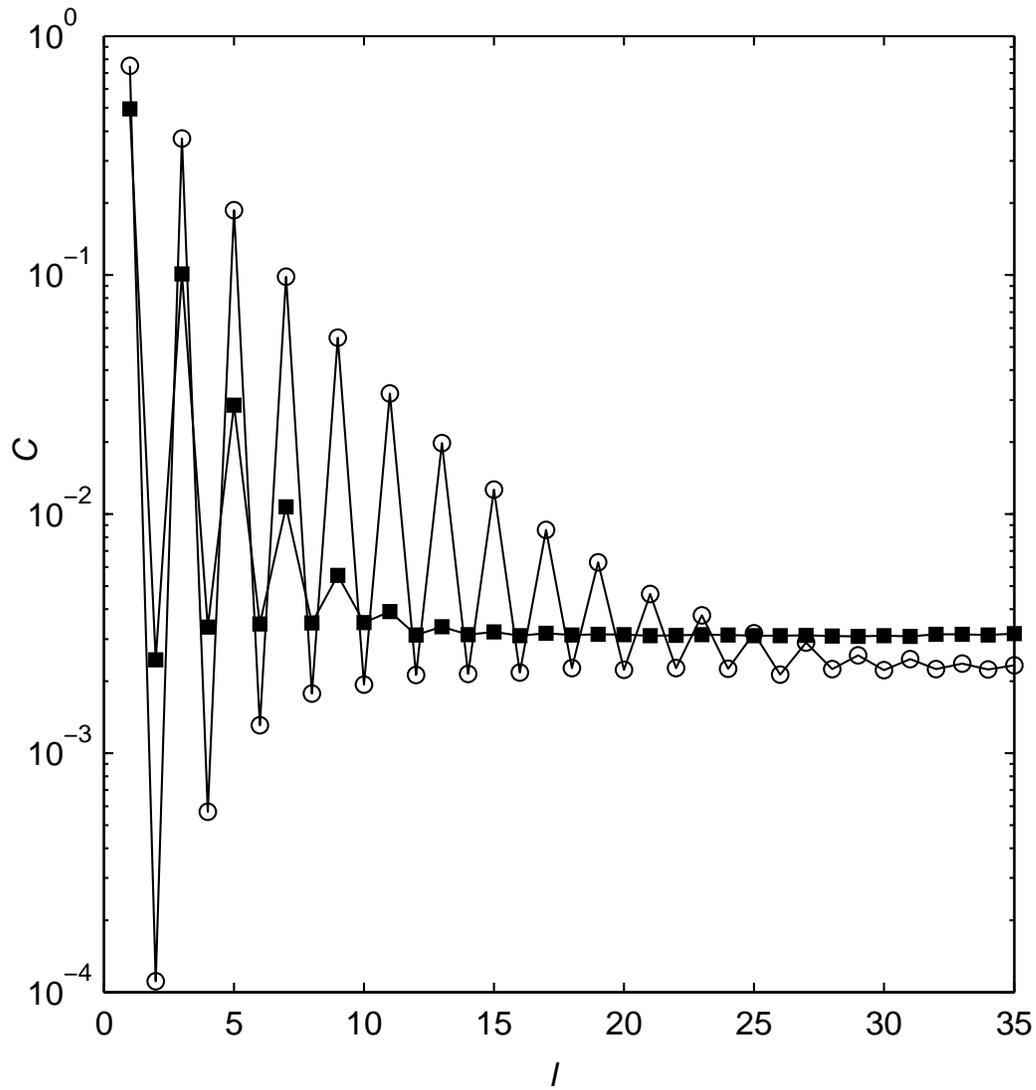}
\end{center}
\caption{Clustering coefficient of random-walk-generated networks as
function of walk length $l$ for $m=2$ ($\circ$) and $m=4$ ($\blacksquare$), with
 $N=25,000$, averaged over 100 runs. 
} \label{fig2}
\end{figure}

\clearpage
\begin{figure}[t]
\begin{center}
\includegraphics[width=\textwidth]{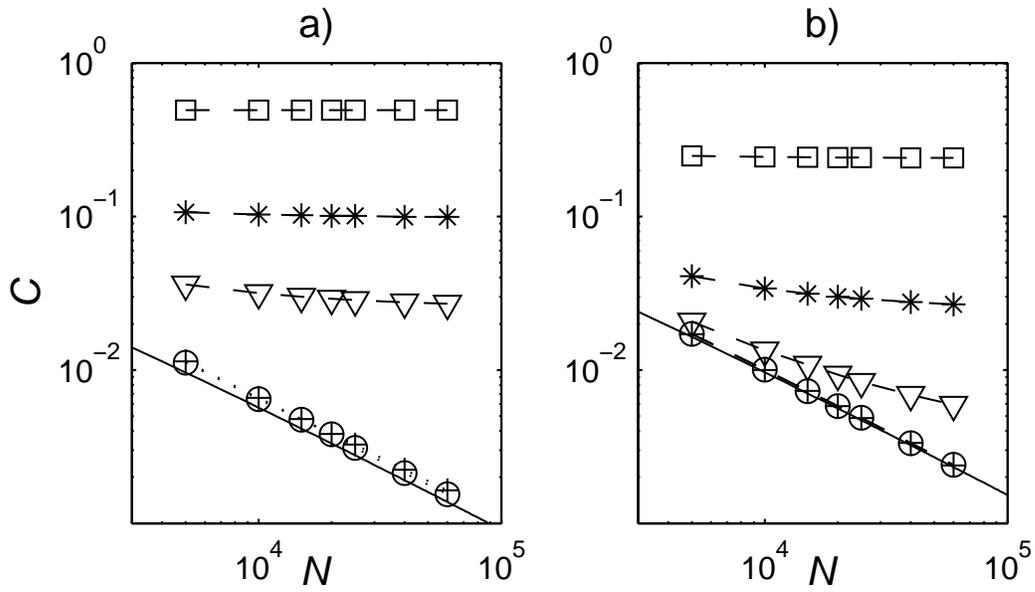}
\end{center}
\caption{Clustering coefficient $C$ as function of network size $N$ for 
a) $m=4$,
b) $m=8$, with random walk lengths $l=1$ ($\square$), $l=3$ ($\ast$),
$l=5$ ($\triangledown$), $l=15$ (+), and $l=25$ ($\circ$), calculated
as averages over 100 simulated runs. The dashed lines serve as guides to
the eye and the solid line displays
the theoretical prediction \cite{Fronczak2003} for clustering coefficients of 
Barab\'{a}si-Albert-networks.\label{fig3}
}

\end{figure}

\end{document}